\documentclass[useAMS,usenatbib]{mn2e}
\pdfoutput=1
\usepackage{graphicx}

\begin{document}

\title[The $CT_1$ Washington system]{New age-metallicity diagnostic diagram for the
Washington photometric system}

\author[A. E. Piatti and G.I. Perren]{Andr\'es E. Piatti$^{1,2}$\thanks{E-mail: 
andres@oac.uncor.edu}, Gabriel I. Perren$^{2,3}$ \\
$^1$Observatorio Astron\'omico, Universidad Nacional de C\'ordoba, Laprida 854, 5000, 
C\'ordoba, Argentina\\
$^2$Consejo Nacional de Investigaciones Cient\'{\i}ficas y T\'ecnicas, Av. Rivadavia 1917, 
C1033AAJ, Buenos Aires, Argentina \\
$^3$Facultad de Ciencias Astron\'omicas y Geof\'{\i}sicas, Universidad Nacional de 
La Plata, Paseo del Bosque s/n, 1900 La Plata, Argentina\\
}

\maketitle

\begin{abstract} 
The age calibration of the Washington $\delta$$T_1$ index is mainly used to estimate ages
of star clusters older than 1 Gyr, no age-metallicity degeneracy effect is considered.
We have profusely exploited synthetic $T_1$ versus $C-T_1$ colour magnitude diagrams 
aiming at exploring the intrinsic behaviour of the $\delta$$T_1$ index. 
The analysis shows that $\delta$$T_1$ varies with age and metal content as well. In general,
the dependence on age weakens for ages greater than $\sim$ 6 Gyr, and
results even less sensitive to age as the metallicity decreases.
For ages younger than $\sim$ 5 Gyr $\delta$$T_1$ shows a strong correlation with
both age and metallicity. The $\delta$$C$ index -defined as $\delta$$T_1$
for the $C$ passband- is also a combined measurement of age and metallicity.
We  introduce a new age-metallicity diagnostic diagram, $\delta$$T_1$ versus
$\delta$$C$ - $\delta$$T_1$, which has shown the ability of unambiguously providing 
age and metallicity estimates,  simultaneously.
 The new procedure allows to derive ages from 1 up to 13 Gyr and 
metallicities [Fe/H] from -2.0 up to +0.5 dex, 
and is independent of the cluster reddening and distance modulus. It does solve the
constraints found in the $\delta$$T_1$ index and surpasses the performance of the
standard giant branch metallicity method. All these features make the diagnostic diagram 
 a powerful tool for 
estimating accurate ages as well as metallicities.

\end{abstract}

\begin{keywords}
techniques: photometric 
\end{keywords}

\section{Introduction}

\citet[][hereafter G97]{getal97} provided a calibration for the magnitude difference
($\delta$) between the giant branch clump in intermediate-age star clusters (the horizontal
branch in old clusters) and the main sequence turnoff as a function of the cluster age 
for the Washington photometric system. Particularly, they used the $T_1$ versus $C-T_1$
colour magnitude diagram (CMD) (effective wavelengths: $C$ $\sim$ 3900\AA, 
$T_1$ $\sim$ 6300\AA; \citet{c76}) of six known star clusters (ages $\ga$ 1 Gyr) in order 
to measure 
$\delta$$T_1$ and fitted those values with the clusters' ages (see their Fig. 5 and
eq. 4). $\delta$$T_1$ emulated the former $\delta$$V$ index defined by 
\citet{petal94}. 
By using such a calibration G97 firstly searched for old star clusters in the
Large Magellanic Cloud; the 25 candidates analysed resulted to be of intermediate age
(1-3 Gyr). Since then, the $\delta$$T_1$ index was employed successively to estimate the
ages of star clusters in the Milky Way \citep[e.g.][]{petal04}, in both Magellanic Clouds 
\citep[e.g.][]{betal98,petal01} and to derive ages for the so-called representative 
stellar population of galactic star fields \citep[e.g.][]{pietal12}, among others.

The age calibration of the $\delta$$T_1$ index is based on $CT_1$ photometry of NGC\,2213 and 
ESO\,121-SC03 in 
the Large Magellanic Cloud and the Galactic open clusters NGC\,1245, Tombaugh\,2, M\,67 and 
NGC\,6791.  At the very least the calibration needs revision,
since updated ages and metallicities for the calibration clusters are available. For 
instance, \cite{atetal07} derived an age of (7.0 $\pm$ 1.0) Gyr for NGC\,6791
([Fe/H]=(+0.42$\pm$0.05) dex Heiter et al. 2014), whereas a value of 10 Gyr was used by G97. 
Except for ESO\,121-SC03 ([Fe/H]=(-0.93$\pm$0.20) dex Olszewski et al. 1991), 
all of these calibration clusters are considerably more metal-rich than [Fe/H] $\sim$ -0.5 dex.
In addition, the solar abundance cluster M\,67 plays a pivotal role in the calibration as it is 
the  only cluster in the 2-9 Gyr range.  
Secondly, thinking in the impact that the adopted cluster fundamental parameters can have 
on the former age calibration of the $\delta$$T_1$ index, it seems necessary to
enlarge the cluster sample in order to support a robust relationship and to
allow an analysis about the possible metallicity sensitivity and the effect any such 
sensitivity might have on the derived ages. Nowadays, however, a representative 
sample of clusters
with age and metallicity values distributed well along the known cluster age/metallicity regime
and with Washington  $CT_1$ photometry is not available.
Given the usefulness that $\delta$$T_1$  has shown in the 
literature as a cluster age indicator, the above constraints seriously blur its full scope.

Fortunately, synthetic cluster CMDs are powerful tools to probe the genuine performance of 
$\delta$$T_1$ as an age indicator and to disentangle any metallicity dependence. This is because 
they can accurately reproduce CMDs of clusters of any age and metallicity, bearing in mind
the fraction of binaries, the cluster initial mass function, the cluster mass, as well as
photometric uncertainties and completeness effects, etc \citep{poetal12,cvdb14}. In this sense,
synthetic cluster CMDs have the advantage of representing those of calibration clusters with 
ages and metallicities derived on a homogeneous scale. 
Indeed, we have taken advantage of the recently developed Automated Stellar Cluster Analysis 
package \citep[\texttt{ASteCA}\footnote{http://asteca.github.io/}][]{petal15}, as we describe 
in Sect. 2 of this work, to profusely exploit a 
large number 
of synthetic cluster CMDs. From the outcomes of this comprehensive analysis, we present in Sect. 
3 a new age-metallicity diagnostic diagram that involves $\delta$ values for the $C$ and 
$T_1$ bandpasses. The proposed technique allows to estimate ages and metallicities for clusters 
older than 1 Gyr, independent of their reddening and distance moduli.
Sect. 4 analyses the performance of the new procedure in the light of  published accurate age and
metallicity values as well as of different age and metallicity calibration for the Washington 
system. Finally, Sect. 5 summarises our main results.
 
\section{Synthetic Colour-Magnitude Diagrams}

Synthetic stellar populations have usually been generated to study the astrophysical properties
of different stellar populations through the comparison of their respective CMDs 
\citep[e.g.][]{retal89,hetal99,metal10}. The recovery of the star formation history
of galaxies \citep{retal12}, the study of the extended main sequence turnoff phenomenon in
star clusters \citep{cetal14}, and the estimation of star cluster parameters \citep{detal15} 
have been topics of astrophysical interest approached by synthetic CMD analyses. 
Likewise, synthetic CMDs offer the possibility to explore in detail the behaviour of different
photometric indices in terms of astrophysical quantities
\citep{catelanetal14}, to predict the range of distinct fundamental properties 
in star clusters \citep{popescuetal14} as well as to assess in advance the performance of 
astronomical instruments \citep{ketal09}. 

In order to revise the $\delta$$T_1$ age calibration and its possible dependence with the
metal content, we have employed the \texttt{ASteCA} suit of functions to generate synthetic CMDs
of star clusters covering ages from 1.0 up to 12.6 Gyr and metallicities in the range
[Fe/H] = (-2.0 - +0.5) dex. \texttt{ASteCA} was designed as a new set of open source tools for an
objective and automatic analysis of large cluster data sets. The code includes
functions to perform cluster structure analysis, luminosity function curves, integrated
colour estimates statistically cleaned from field star contamination, a Bayesian membership 
assignment algorithm, and a synthetic cluster-based best isochrone matching method to 
simultaneously estimate clusters' properties (age, metallicity, distance, reddening, mass and binarity).
\citet{petal15} showed that it does not introduce any biases or new correlations
between the various derived cluster parameter values.

The steps by which a synthetic cluster for a given set of age, [Fe/H], distance modulus $m-M$, and 
reddening $E(B-V)$ values is generated by \texttt{ASteCA} is as follows: i) a theoretical isochrone
is picked up, densely interpolated to contain a thousand points throughout its entire length,
including the most evolved stellar phases. ii) The isochrone is shifted in colour and
magnitude according to the $E(B-V)$ and $m-M$ values to emulate the effects these extrinsic 
parameters have over the isochrone in the CMD. At this stage the synthetic cluster 
can be objectively identified as a unique point in the 4-dimensional space of 
parameters ($E(B-V)$, $m-M$, age and metallicity). iii) The isochrone is trimmed down to a certain faintest magnitude 
according to the limiting magnitude thought to be reached. iv) An initial mass function 
(IMF) is sampled in the mass range $[{\sim}0.01{-}100]\,M_{\odot}$ up
to a total mass value $M_{total}$ provided via an input data file that
ensures the evolved CMD regions result properly populated.
Currently, \texttt{ASteCA} lets the user choose 
between three IMFs \citep{Kroupa_1993,Chabrier_2001,Kroupa_2002} but there is 
no limit in the number of distinct IMFs that can be added. 
The distribution of masses is then used to obtain a properly populated synthetic 
cluster by keeping one star in the interpolated 
isochrone for each mass value in the distribution. v) A random fraction of stars are 
assumed to be binaries, which is set by default to  
$50\%$ \citep{von_Hippel_2005}, with secondary masses 
drawn from a uniform distribution between the mass of the primary star and a 
fraction of it given by a mass ratio parameter set to $0.7$. Both 
quantities can be modified through the input data file. vi) An appropriate
magnitude completeness and an exponential photometric error functions are
finally applied to the synthetic cluster. 

As for our purposes, we used the theoretical isochrones computed by \citet{betal12}
using extensive tabulations of bolometric corrections with uncertainties $\sim$ 0.001 mag
for the $C$ and $T_1$ filters, the
IMF of \citet{Chabrier_2001}, and a cluster mass as a function of the cluster's age
given by the expression  log($M$$_{total}\,\,\,   $M$_{\odot}$) = 
1.8$\times$log(age yr$^{-1}$) - 12.8 \citep{baetal13,dgetal13,p14b} in order to keep not 
only the main sequence turnoff (MSTO) 
but also the red clump (RC) similarly populated as those of known low mass clusters.
We considered no binarity effect, since it has been shown
that binary stars can broaden the MSTO region \citep{letal12,letal13,jetal14}, in contrast
with our aim of sizing up the sensitivity of the MSTO to age and metallicity. 
Fig. 1 shows some examples of the resulting synthetic CMDs. 
 Note that the theoretical isochrones of \citet{betal12} are in a very satisfactory 
agreement with another known set of isochrones for the Washington system computed
by \citet{lsh01} \citep[e.g.][and references therein]{petal03b}.

\section{The $\delta$$T_1$ index}

 We measured the difference in magnitude between the MSTO and the RC in the
generated synthetic CMDs 
following the precepts of \citet[][see their Fig. 1]{petal94}, as G97 also applied
in their definition of $\delta$$T_1$. The synthetic cluster CMDs allowed us to estimate the 
errors involved in measuring $\delta$$T_1$. We found that an uncertainty between 0.05-0.10 mag
typically affects the measurements of $\delta$$T_1$, and can reach a few more hundredth of 
magnitude for some of the youngest clusters in our synthetic sample because of the shape of 
their MSTOs.
Fig. 2 (upper left-hand panel) shows the resulting $\delta$$T_1$ values as a function
of age for six different metallicity levels (coloured line scale). As can be seen, 
the relationship of $\delta$$T_1$ with age varies with the metal content as well.
In general, the trend of $\delta$$T_1$ with the age weakens (it changes to a smaller slope) 
for ages larger than $\sim$ 6 Gyr, and results even less sensitive to age as
the stellar population becomes more metal-poor ([Fe/H] $\la$ -1.0 dex).  This means that 
$\delta$$T_1$ does not result a good indicator of age independent of metallicity for old 
and metal-poor clusters. For ages younger than $\sim$ 5 Gyr $\delta$$T_1$ shows a strong 
correlation with both age and metallicity. 

We included in Fig. 2 (upper left-hand panel) the relationship obtained by G97 represented
by a solid black line. For ages younger than 3 Gyr they used three calibration clusters,
two of them (NGC\,2243, Tombaugh\,2) have an average spectroscopic metallicity of 
[Fe/H] = (-0.4 $\pm$ 0.10) dex. Their $\delta$$T_1$ versus age relationship 
roughly matches the synthetic one for [Fe/H] = -0.5 dex. On the other hand, the older end of the G97's 
curve overlaps those of metal-poor clusters ([Fe/H] $\la$ -1.5 dex). The upper right-hand panel
of Fig. 2 illustrates clearer this metallicity dependence of $\delta$$T_1$ for twelve 
different age levels. 
Indeed, for 
$\delta$$T_1$ = 3.2 mag an age of 10.3 Gyr comes out from the G97's calibration as well as from 
synthetic CMDs with [Fe/H] $\la$ -1.5 dex. However, an age $\sim$ 2.5 Gyr smaller results from
the synthetic CMDs if a [Fe/H] $>$ -1.0 dex
is adopted. Such a difference between mean age values results statistically
significant even if we consider an uncertainty of 0.15 mag in the $\delta$$T_1$ values.
Similarly, for $\delta$$T_1$ = 1.5 mag an age of 1.8 Gyr is derived from the G97's equation,
or indistinctly from synthetic CMDs with [Fe/H] $\approx$ -0.6 dex. Once again, if [Fe/H] 
values of -0.1 dex and -1.1 dex were instead taken into account, the ages from synthetic CMDs 
would result 0.5 Gyr 
greater and 0.6 Gyr smaller than the G97's age, respectively; errors of 0.2 Gyr due to 
uncertainties in $\delta$$T_1$ were estimated.

The above examples not only illustrate that the $\delta$$T_1$ age index is sensitive
to metallicity, but also that such a dependence is a complex function of both
age and metallicity. Moreover, the age range for any particular $\delta$$T_1$ value
obtained from the synthetic curves (see upper left-hand panel of Fig. 2) results many times 
larger than the age errors 
derived from eq. 4 of G97. Fig. 3 illustrates this trend for three
different $\delta$$T_1$ error levels.
This result points out to the need of a new calibration for the $\delta$$T_1$ age 
index which prevents us against any age/metallicity degeneracy. 

We seek for any possible straightforward relationship between the $\delta$$T_1$ values 
derived from synthetic cluster CMDs, the age and the metallicity. Bearing in mind 
the kind of arithmetic expressions employed in previous $\delta$ age calibrations 
\citep{petal94,cc94,getal97,setal04}, we tried different possibilities which
included linear, quadratic and terms of higher degree in any of the three
quantities, mixed terms, logarithmic functions, etc. Note that we used 
a larger number of points than any previous $\delta$ age calibration, uniformly
distributed throughout the whole age/metallicity range and without any constraint
from non-homogeneity in the age/metallicity values. Unfortunately, we did not
attain any satisfactory fit, which confirms the complex interdependence of the three
parameters. 

In order to complement
this analysis and by taking advantage of the availability of magnitudes in the $C$
passband (= $T_1$ + ($C-T_1$)), Fig. 2 (bottom panels) also depicts
the relationship for $\delta$$C$ - defined as $\delta$$T_1$ but for the $C$ filter -  
with the age and the metallicity. In order to build that relationship, we first
produced synthetic $C$ versus $C-T_1$ CMDs for the same synthetic clusters used to 
construct the top panels of Fig. 2. Then, we
measured the magnitudes at the MSTO and the RC and computed their difference 
($\delta$$C$). The resulting curves show that $\delta$$C$ expands
over a dynamical range of $\sim$ 4 mag, similar to that of $\delta$$T_1$. 
However, although the overall appearance of $\delta$$C$ and $\delta$$T_1$ curves look 
alike, the former show a less complex trend with age and
metallicity. Particularly, they do not account for: i) the lack of metallicity
sensitivity ([Fe/H] $>$ -1.0 dex) for stellar populations older than $\sim$ 5 Gyr
and, ii) the slope change for clusters more metal-poor than [Fe/H] $\la$ -1.0 dex.
Nevertheless, Fig. 2 clearly suggests that $\delta$$C$ is also a combined measurement 
of age and metallicity.

\subsection{A new age-metallicity diagnostic diagram}

At this point, we decided to introduce a new diagnostic diagram with the ability
of unambiguously providing age and metallicity estimates within certain
 Washington $CT_1$ photometric error limits. We found that the plane $\delta$$T_1$
versus $\delta$$C$ - $\delta$$T_1$ resulted to be the one which best distinguishes
changes in age and metallicity throughout the whole two-dimensional space. In
Fig. 4 we have traced iso-age lines and marked iso-abundance positions using
colour-coded lines and filled circles, respectively. Errobars for typical 
uncertainties in $\delta$$T_1$ and $\delta$$C$ - $\delta$$T_1$ are also indicated. 
Note that the iso-age lines and iso-metallicity positions 
rely on the theoretical isochrones used to build
the synthetic CMDs, while the quoted errors refer to the uncertainty in measuring 
$\delta$$C$ and $\delta$$T_1$ in observed
cluster CMDs. In this sense, the capacity of resolving ages and
metallicity varies with the position in that plane. Table 1 lists
the age/metallicity errors associated to typical  $\sigma$($\delta$$T_1$) and
$\sigma$($\delta$$C$ - $\delta$$T_1$) uncertainties,  derived by interpolation
of Fig. 4. Particularly, we have highlighted 
with boldface characters those  errors for (age, [Fe/H]) pairs that are within
the ranges of the age-metallicity relationships (AMRs) of the Milky Way and of the 
Small/Large Magellanic Clouds \citep{beetal04,setal09,pg13}.  Since star
clusters can have ages and metallicities directly linked to the chemical evolution 
history (AMR) of their host galaxies, the highlighted age/metallicity ranges in 
Table 1 result more usable from an astrophysical point of view.
We thickened the line sections in the $\delta$$T_1$ versus age diagram (see
Fig. 5) corresponding to age/metallicity values with uncertainties in Table 1 
in boldface. Thus,  readers can consistently compare the G97 locus with the present 
theoretically-driven calibration according to known AMRs.
Fig. 4 is aimed at entering with
($\delta$$C$-$\delta$$T_1$, $\delta$$T_1$) values and obtaining by interpolation
age and metallicity estimates.

We used the new diagnostic diagram for clusters with high-quality
$CT_1$ photometry, particularly with well-identified MSTOs and RCs,  and 
with accurate ages and metallicities. We searched for metallicities obtained from
high-dispersion spectroscopy, although in some cases we relaxed this requirement
down to medium-dispersion spectroscopy or even to reliable photometric
metal abundances. As for the cluster ages, we took advantages of those values
derived from isochrones fitting to deep cluster CMDs.
We excluded any previous age/metallicity estimates coming from Washington
photometry. 

From barely 340 clusters with published $CT_1$ data, we found nearly 150 with
recognisable MSTOs and RCs. In general, the extracted cluster CMDs 
present signatures of contamination from the composite star field
population along the line of sight. Such a field contamination does have
a particular pattern given by the luminosity function, the colour distribution 
and the stellar density towards the cluster region. For these reasons, we first 
built CMDs representing
the field along the line of sight towards the
individual clusters, which we then used to clean the cluster CMDs with the aim
of tracing fiducial cluster sequences as accurate as possible.
We employed the cleaning procedure developed by \citet[see their
Fig. 12]{pb12}. The method compares the extracted cluster CMD to
four distinct CMDs composed of stars located reasonably far from the
object, but not too far so as to risk losing the local field-star
signature in terms of stellar density, luminosity function and/or
colour distribution. Each field region covers an
equal area as that of the cluster and the four field areas are placed to the north, 
east, south and west from the cluster.

The comparison of the cluster/field CMDs is performed using 
boxes of different sizes distributed in
the same manner throughout both CMDs, thus leading to a more meaningful
comparison of the numbers of stars in different CMD regions than
using boxes fixed in size and position. The latter is not universally efficient,
since some parts of the CMD are more
densely populated than others. For instance, to
deal with stochastic effects at relatively bright magnitudes (e.g.,
fluctuations in the numbers of bright stars), larger boxes are
required, while populous CMD regions can be characterized using
smaller boxes.  Since the procedure is executed for each of the four 
field-CMD box samples, it assigns a membership probability (P) to each 
star in the cluster CMD. This is done by counting the number of times a 
star remained unsubtracted in the four cleaning runs and by
subsequently dividing this number by four. For our purposes, 
we used stars that are predominantly found in the cleaned cluster
CMDs ($P \ge$ 75\%).  

We used the cleaned cluster CMDs to measure $C$ and $T_1$ magnitudes at the
MSTO and RC, then computed $\delta$$C$ and $\delta$$T_1$ and entered into
the age-metallicity diagnostic diagram to estimate cluster ages and metallicities.
The resulting values with their uncertainties for clusters that fulfilled  the 
requirements mentioned above (e.g. available age/metallicity values from independent
Washington techniques) are listed in the last columns
of Table 2, in which we also included age/metallicity values adopted by thoroughly 
searching the literature for comparison purposes.

\section{Analysis}

From the comparison between cluster ages taken from the literature and those estimated
above (see Fig. 6) we obtained a mean age dispersion of $\Delta$(log(age yr$^{-1}$) =
0.07 $\pm$ 0.02 along the whole age range (1 - 13 Gyr). The resulting mean dispersion is 
slightly smaller than typical age errors ($\Delta$(log(age yr$^{-1}$) = 0.10-0.15) derived from 
isochrone fitting to good-quality cluster CMDs, regarless of binarity, multiple 
populations, differential reddening, rotation effects, etc. This result shows that
the age-metallicity diagnostic diagram returns accurate ages at low expense, since
it does not require deep photometry nor deal with the known 4-parameter degeneracy 
(age, metallicity, distance, reddening) when matching isochrones to the cluster CMDs.
On the metallicity arena, the derived iron to hydrogen ratios are within $\Delta$([Fe/H])
= $\pm$ 0.15 dex of the identity relation (see Fig. 7), no systematic dependence with
the metallicity taken from the literature is visible along the considered range.
Such a dispersion is also comparable to the smallest error atainable in deriving
metal abundances from photometric data.  The most discrepant point in Fig. 7 corresponds
to NGC\,6791, which is simply due to a drop in the metallicity sensitivity for
metal-rich clusters older than 5 Gyr.

These results reveal that the diagnostic diagram is able to solve the constrains
found in the $\delta$$T_1$ index, namely, the loss of age sensitivity for ages greater
than $\sim$ 6 Gyr and the strong dependence on both age and metallicity for ages smaller
than $\sim$ 5 Gyr. Indeed, as for metal-poor and old clusters, the loss of age sensitivity
in $\delta$$T_1$ is surpassed when the $\delta$$C$ - $\delta$$T_1$ index is used 
as a variable instead of the age as proposed by G97 (see top left-hand panel of Fig. 2). 
Such a choice also allows to obtain a metallicity estimate. For ages younger than $\sim$ 5 Gyr,
the diagnostic diagram shows that while $\delta$$T_1$ depends on the age and on the metallicity,
$\delta$$C$ - $\delta$$T_1$ is mainly a metal abundance indicator, so that the latter
fixes the metallicity level where the former is evaluated.

In addition to the $\delta$$T_1$ age calibration, the $M_{T_1}$ versus $(C-T_1)_o$ CMD
was also calibrated in terms of metallicity by \citet{gs99}, who demonstrated
the metallicity sensitivity of the standard giant branch (SGB, each giant branch corresponds 
to an iso-abundance curve) applicable to objects with ages $\ga$ 5 Gyr, any age 
effects are small or negligible for such objects.
However,  the SGBs were defined for [Fe/H] $<$ -0.5 dex using globular clusters older than 
10 Gyr, so that it is important to examine as closely as
possible the effect of applying such a calibration  to much younger clusters. 
In view of the
well-known age-metallicity degeneracy, \citet{betal98} explored this effect empirically
by comparing SGB-based metallicities for 11 clusters with ages between 1 and 3 Gyr
to standard values.  They found a relatively constant offset of $\sim 0.4$ dex,
the SGB metallicities were underestimated due to the effects
of age for clusters younger than $\sim$ 3 Gyr. \cite{getal03} investigated
this effect in much more detail by using theoretical isochrones 
computed by \citet{lsh01} for two metallicity levels ([Fe/H] =
-1.3 dex and -0.7 dex).  They found that not only a constant offset of
$\sim 0.4$ dex but an
exponential correction increasing towards younger ages is necessary.
Particularly, they adopted the theoretical prediction for [Fe/H] = -0.7 dex as
the correction to be 
applied to the SGB metallicities as a function of age. Including all error sources,
the corrected [Fe/H] values are estimated with an uncertainty of $\sigma$([Fe/H]) = 0.3 dex,
although the steepness of the age correction for the 
youngest clusters ($<$ 2 Gyr) results in a larger metallicity
error and bias the resulting metallicities upwards.

The referred successive improvements seem surpassed by the new age-metallicity
diagnostic diagram. The latter can be used without the knowledge of the cluster 
distance and reddening, which is mandatory in the SGB technique. Likewise, the
diagnostic diagram allows not only to estimate directly more precise metallicities 
but also  ages, simultaneously. The new procedure is subject of no correction
and useful for a wide range of ages (1 - 13 Gyr) and metallicities ([Fe/H] = -2.0 - 
+0.5 dex).  All these feature make the diagnostic diagram a powerful tool 
for estimating accurate ages as well as metallicities.

\section{Conclusions}

The Washington photometric system has long been used to estimate ages and metallicities
of clusters, particularly for those older than $\sim$ 1 Gyr. Nevertheless,
these estimates have relied on calibrations ($\delta$$T_1$ and SGB methods) which involve
clusters with ages and metallicities that need to be updated. Likewise, the
well-known age-metallicity degeneracy has not been properly addressed or even
not considered at all. 

We have profusely exploited synthetic $T_1$ versus $C-T_1$ CMDs with the aim of improving 
our knowledge about the intrinsic behaviour of the $\delta$$T_1$
index with age and metallicity. The synthetic CMDs were produced through the
\texttt{ASteCA} suit of functions, taking into account the total cluster mass as a function
of age in order to have the MSTO and the RC similarly populated to those known low mass
clusters. Photometric errors were also considered, so that the resulting cluster CMDs
achieved the appearance of the observed ones.

The analysis of the $\delta$$T_1$ index as a function of the age for different
metallicity levels shows that it varies with age and metal content as well. In general,
the dependence on age weakens for ages greater than $\sim$ 6 Gyr, and
results even less sensitive to age as the metallicity decreases ([Fe/H] $\la$ -1.0 dex).
For ages younger than $\sim$ 5 Gyr $\delta$$T_1$ shows a strong correlation with
both age and metallicity. As expected, the $\delta$$C$ index -defined as $\delta$$T_1$
for $C$ the passband - is also a combined measurement of age and metallicity.

We  introduce a new age-metallicity diagnostic diagram, $\delta$$T_1$ versus
$\delta$$C$ - $\delta$$T_1$, which has shown the ability of unambiguously providing 
age and metallicity estimates, simultaneously, within certain Washington $CT_1$  photometric 
error limits.  The proposed technique does not require any additional measurement
from other Washington passbands, but only the same $CT_1$ photometry needed to
measure the former $\delta$$T_1$ index. The new procedure allows to derive ages and 
metallicities within a considerable wide range (age: 1 - 13 Gyr, [Fe/H]: -2.0 - +0.5 dex), 
and is independent of the cluster reddening and distance modulus. It does solve the
constraints found in the $\delta$$T_1$ index and surpasses the performance of the
SGB method.

\section*{Acknowledgements}
This work was partially supported by the Argentinian institutions
CONICET and Agencia Nacional de Promoci\'on Cient\'{\i}fica y
Tecnol\'ogica (ANPCyT).
We are grateful for the comments and suggestions raised by the anonymous
referee which helped us to improve the manuscript.

\bibliographystyle{mn2e_new} 
\bibliography{paper} 
%
%

\clearpage

\begin{table*}
\caption{Estimated age (Gyr) and metallicity (dex) errors$^*$}
\begin{tabular}{@{}lcccccc}\hline
{\rm [Fe/H]}/ age & 1.0 & 2.0 & 3.2 & 4.0 & 5.0 & 6.3 \\\hline
-2.0    &  0.20/0.15 &  0.20/0.20  &  0.30/0.20 &   0.50/0.25 &   0.70/0.20 &   0.80/0.25\\
-1.5    &  0.20/0.15 &  0.20/0.15  &  0.25/0.20 &   0.50/0.25 &   {\bf 0.60/0.25} &   {\bf 0.70/0.25}\\
-1.0    &  {\bf 0.20/0.15} &  {\bf 0.30/0.15}  &  {\bf 0.40/0.25} &   {\bf 0.60/0.25} &   {\bf 0.60/0.25} &   {\bf 0.60/0.25}\\
-0.5    &  {\bf 0.20/0.15} &  {\bf 0.25/0.15}  &  {\bf 0.40/0.25} &   {\bf 0.50/0.20} &   {\bf 0.50/0.25} &   {\bf 0.50/0.25}\\
0.0     &  {\bf 0.15/0.20} &  {\bf 0.15/0.20}  &  {\bf 0.20/0.20} &   {\bf 0.40/0.20} &   0.40/0.20 &   0.50/0.40\\
+0.5    &  {\bf 0.10/0.20} &  0.10/0.20  &  0.20/0.20 &   0.30/0.20 &   0.30/0.20 &   0.40/0.50 \\\hline 
{\rm [Fe/H]}/ age        & 7.1        & 8.0           & 9.0          &  10.0         &  11.2       &  12.6 \\\hline
-2.0    &  0.60/0.20 &  0.60/0.15  &  0.70/0.15 &   0.70/0.15 &   {\bf 0.70/0.15} &   {\bf 0.70/0.15}\\
-1.5    & {\bf 1.00/0.20}  & {\bf 1.00/0.15}   & {\bf 1.00/0.15}  &  {\bf 1.00/0.15}  &  {\bf 1.00/0.15}  &  {\bf 1.00/0.15}  \\ 
-1.0    &  {\bf 0.80/0.25} &  {\bf 1.00/0.25}  &  {\bf 1.00/0.25} &   {\bf 1.00/0.20} &   {\bf 1.00/0.15} &   {\bf 1.00/0.15}\\
-0.5    & {\bf 0.50/0.20}  & {\bf 0.70/0.20}   & {\bf 0.70/0.20}  &  {\bf 0.60/0.20}  &  0.60/0.20  &  0.60/0.20   \\ 
0.0     &  0.50/0.40 &  0.50/0.40  &  0.50/0.40 &   0.50/0.40 &   0.50/0.40 &   0.50/0.40 \\
+0.5    & 0.80/0.50  & 0.80/0.50   & 0.80/0.50  &  0.80/0.50  &  0.80/0.50  &  0.80/0.50\\\hline

\end{tabular}

\noindent $*$ Errors in age/metallicity were estimated using $\sigma$($\delta$$C$) = $\sigma$($\delta$$T_1$) = 0.05 mag and
$\sigma$($\delta$$C$ - $\delta$$T_1$) = [($\sigma$$C$)$^2$ + ($\sigma$$T_1$)$^2$]$^{1/2}$ =
 0.07 mag.  
\end{table*}

\begin{table*}
\caption{Cluster parameters taken from the literature used for comparison purposes.}
\begin{tabular}{@{}lccccccc}\hline
Cluster & Age (Gyr) & Ref. & [Fe/H] (dex) & Ref. & $CT_1$ data & Age (Gyr) & [Fe/H] (dex)  \\
        &           &       &             &      &     Ref.     &   \multicolumn{2}{c}{(this work)} \\\hline
47\,Tuc        & 13.1 $\pm$ 0.9  & 1 & -0.75$\pm$ 0.04 & 11 & 15  & 13.0  $\pm$ 1.0 &-0.80  $\pm$ 0.30\\
AM\,3          & 4.9  $\pm$ 1.8  & 2 &-0.8  $\pm$ 0.4  & 2  & 16  &  4.5  $\pm$ 0.7 &-0.75  $\pm$ 0.40\\
ESO\,121-SC03  & 9.0  $\pm$ 0.7  & 3 &-0.93 $\pm$ 0.20 & 12 & 17  &  9.0  $\pm$ 1.0 &-0.90  $\pm$ 0.30\\
HW\,40         & 2.50 $\pm$ 0.35 & 2 &-0.90 $\pm$ 0.15 & 2  & 16  &  3.0  $\pm$ 0.7 &-1.00  $\pm$ 0.40\\
IC\,2146       & 1.55 $\pm$ 0.05 & 4 &-0.4  $\pm$ 0.2  & 4  & 18  &  1.5  $\pm$ 0.3 &-0.60  $\pm$ 0.25\\
Lindsay\,3     & 1.2  $\pm$ 0.3  & 2 &-0.40 $\pm$ 0.15 & 2  & 19  &  1.3 $\pm$  0.2 &-0.50  $\pm$ 0.15\\
Lindsay\,113   & 4.0  $\pm$ 0.7  & 2 &-1.24 $\pm$ 0.11 & 2  & 20  &  3.5  $\pm$ 0.3 &-1.20  $\pm$ 0.30\\
NGC\,339       & 6.0  $\pm$ 0.5  & 5 &-1.08 $\pm$ 0.12 & 5  & 16  &  5.2 $\pm$  1.0 &-1.00 $\pm$  0.40 \\         
NGC\,419       & 1.4  $\pm$ 0.2  & 6 &-0.67 $\pm$ 0.12 & 13 & 21  &  1.3 $\pm$  0.3 &-0.75  $\pm$ 0.30\\
NGC\,2682      & 4.2  $\pm$ 0.2  & 7 &0.00  $\pm$ 0.06 & 23 & 15  &  4.2  $\pm$ 1.0 &-0.10  $\pm$ 0.30\\
NGC\,6791      & 7.0  $\pm$ 1.0  & 8 & 0.42 $\pm$ 0.05 & 23 & 15  &  7.5  $\pm$ 0.8 & 0.00  $\pm$ 0.50 \\ 
SL\,509        & 1.2  $\pm$ 0.2  & 9 &-0.54 $\pm$ 0.09 & 9  & 17  &  1.0  $\pm$ 0.2 &-0.40 $\pm$  0.15\\
SL\,862        & 1.7  $\pm$ 0.2  & 9 &-0.47 $\pm$ 0.10 & 9  & 17  &  1.7  $\pm$ 0.4 &-0.50  $\pm$ 0.25\\
Trumpler\,5    & 3.0  $\pm$ 1.0  & 10 &-0.40$\pm$ 0.05 & 10 & 22  &  3.0 $\pm$  0.5 &-0.40  $\pm$ 0.15\\\hline

\end{tabular}

\noindent Ref.: (1) \citet{roetal14}; (2) \citet{detal14}; (3) \citet{maetal06};
(4) \citet{mietal09}; (5) \citet{getal11}; (6) \citet{getal09};
(7) \citet{bnetal07}; (8) \citet{atetal07}; (9) \citet{shetal10}; (10) \citet{detal15};
(11) \citet{bretal10}; (12) \citet{oetal91}; (13) \citet{dh98};
(14) \citet{setal92}; (15) \citet{gs99}; (16) \citet{p11c}; (17) \citet{betal98};
(18) \citet{p11a}; (19) \citet{pietal11}; (20) \citet{petal07}; (21) 
\citet{p11b}; (22) \citet{petal04}; (23) \citet{hetal14}.

\end{table*}

\clearpage

\begin{figure*}
\includegraphics[width=144mm]{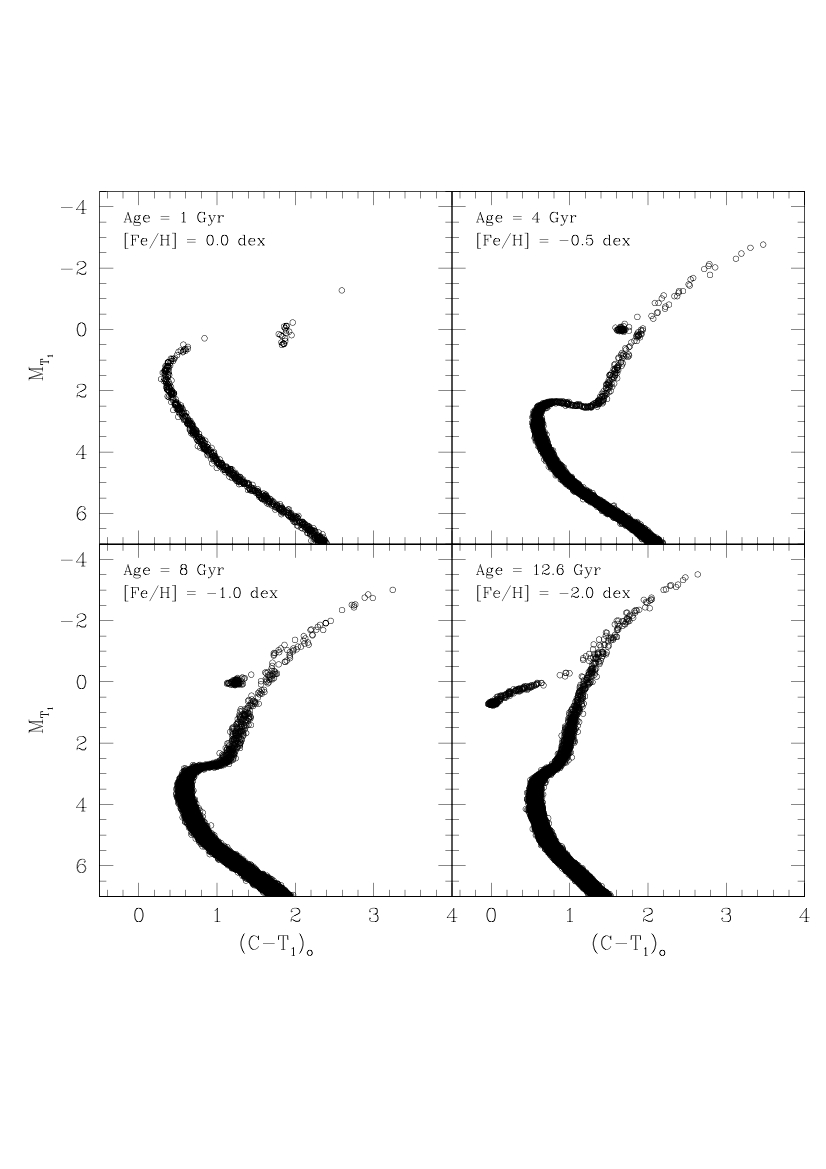}
\caption{Examples of synthetic CMDs produced using the \texttt{ASteCA} suit of functions.}
\label{fig1}
\end{figure*}

\begin{figure*}
\includegraphics[width=144mm]{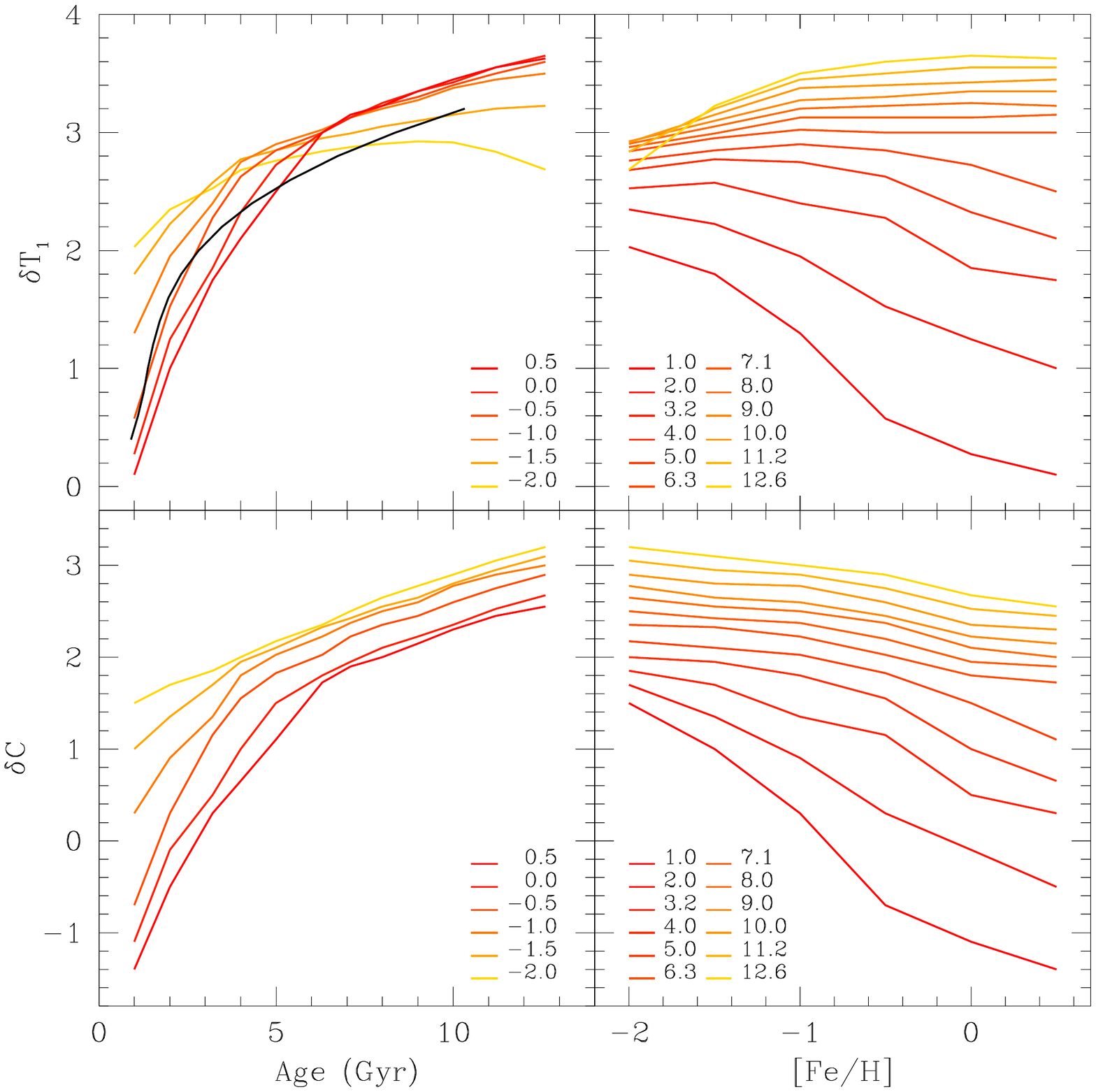}
\caption{Relationship of $\delta$$C$ and $\delta$$T_1$ indices with the age for different
[Fe/H] (dex) values (left-hand panels) and with the metallicity for different ages (Gyr) (right-hand
panels). The black curve in the upper left-hand panel corresponds to the G97's calibration.}
\label{fig2}
\end{figure*}

\begin{figure*}
\includegraphics[width=144mm]{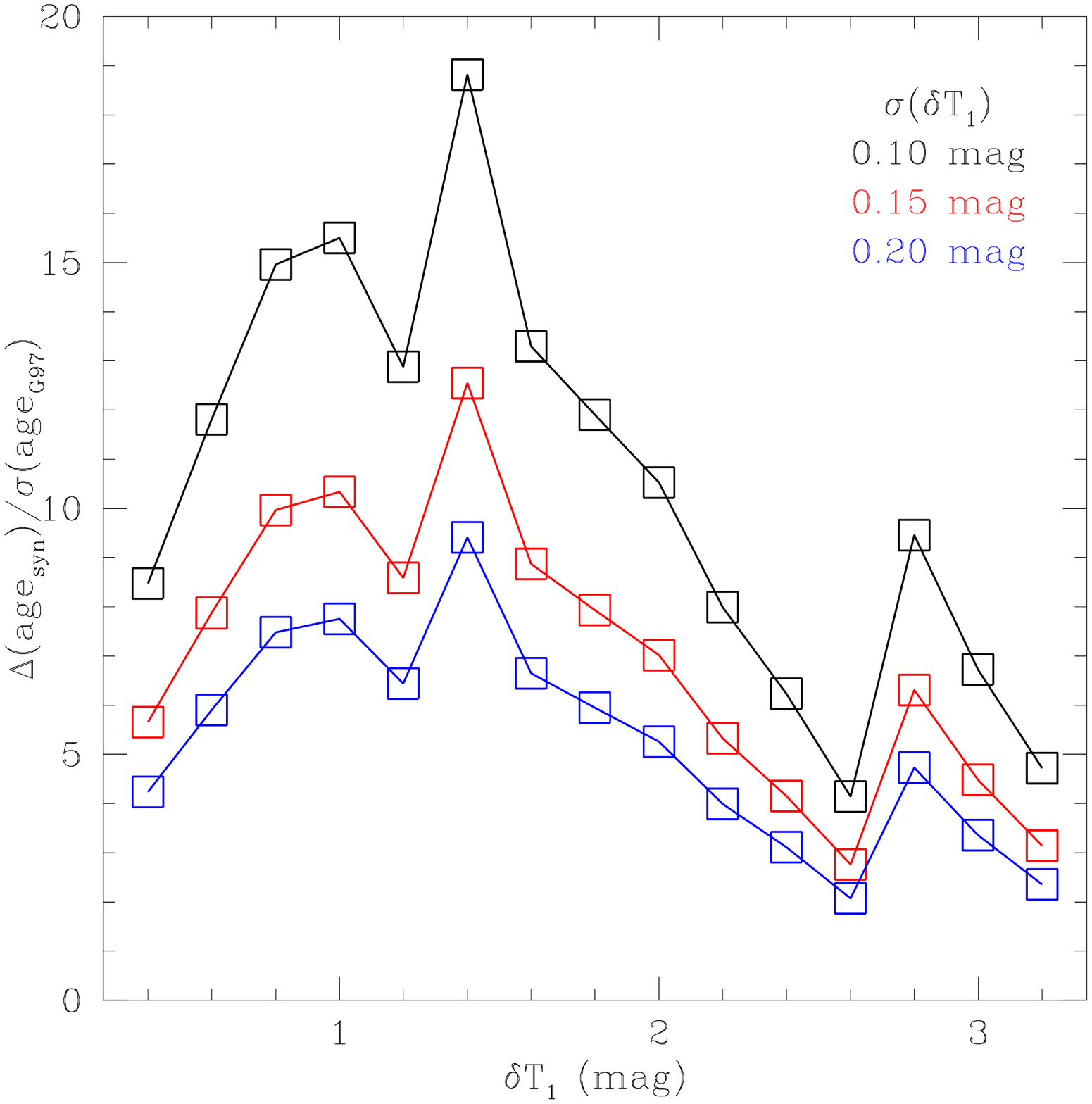}
\caption{Synthetic age range to $\sigma$(age$_{G97}$) ratio as a function of $\delta$$T_1$.}
\label{fig3}
\end{figure*}

\begin{figure*}
\includegraphics[width=144mm]{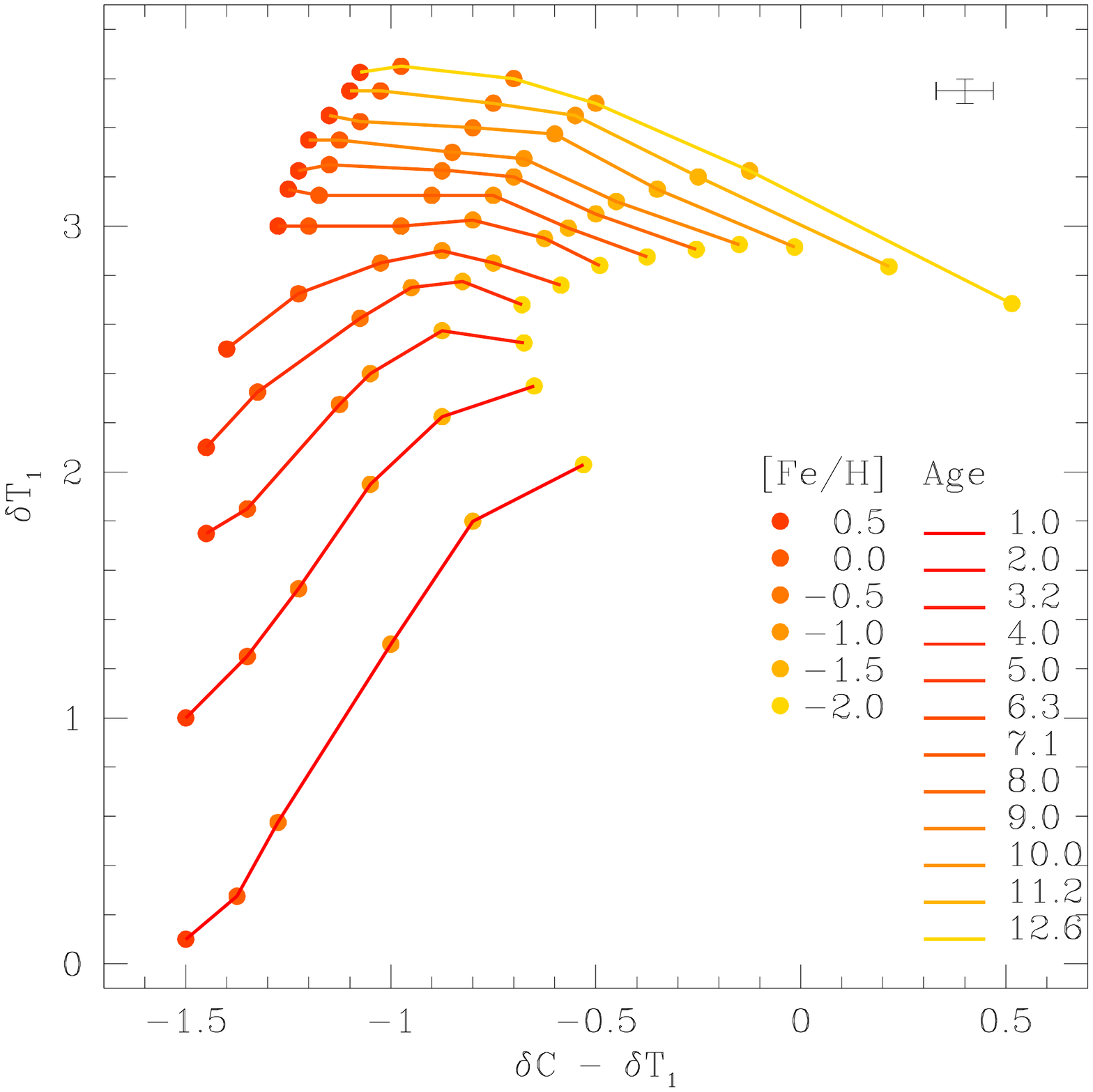}
\caption{$\delta$$T_1$ versus $\delta$$C$ - $\delta$$T_1$ diagram with iso-age lines and 
iso-metallicity locii. Metallicity and age labels are given in dex and Gyr, respectively.}
\label{fig4}
\end{figure*}

\begin{figure*}
\includegraphics[width=144mm]{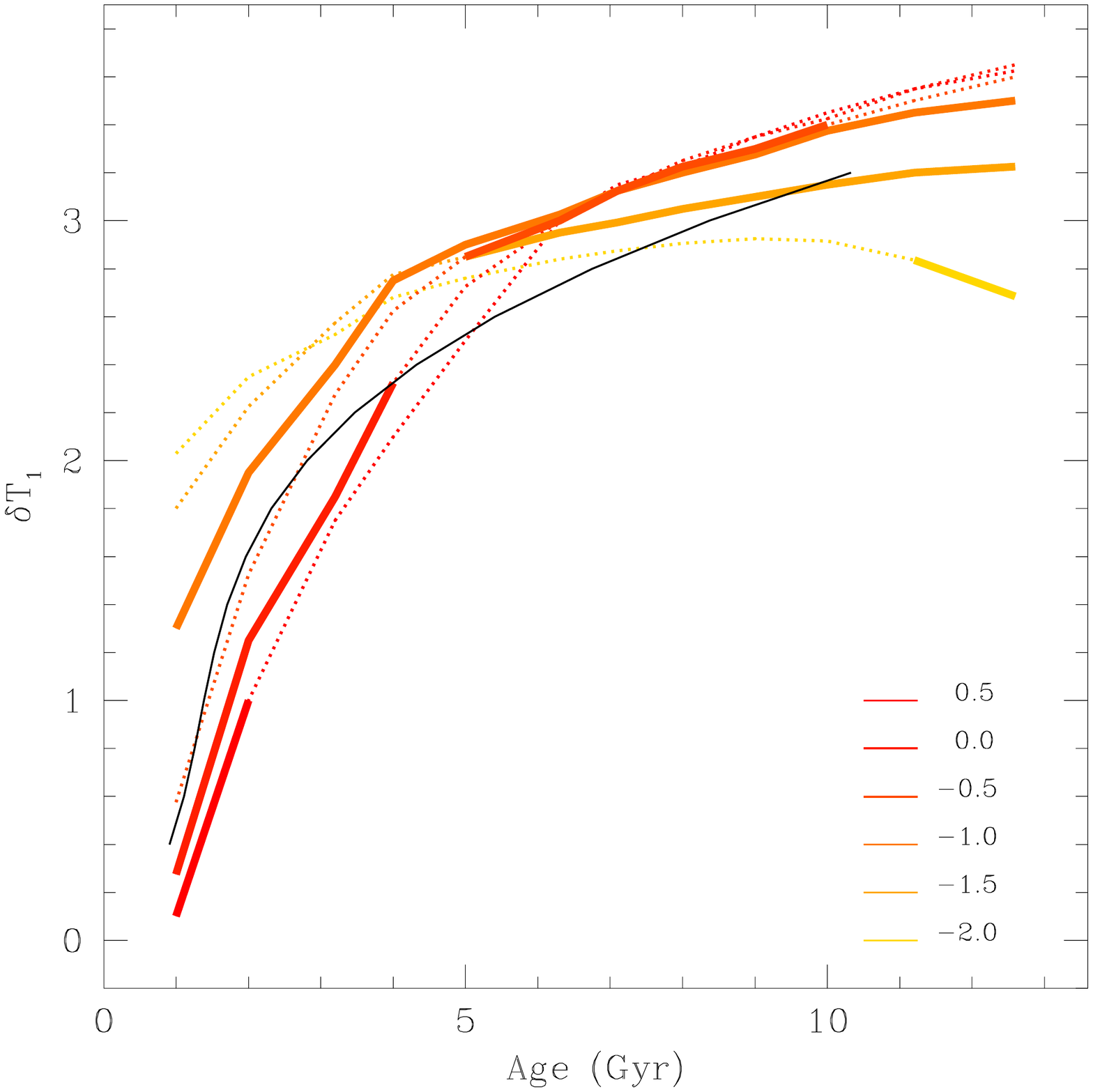}
\caption{Same as Fig. 2 (upper left-hand panel), where thick line sections correpond
to ages/metallicities with uncertainties in boldface characters in Table 1.}
\label{fig5}
\end{figure*}

\begin{figure*}
\includegraphics[width=144mm]{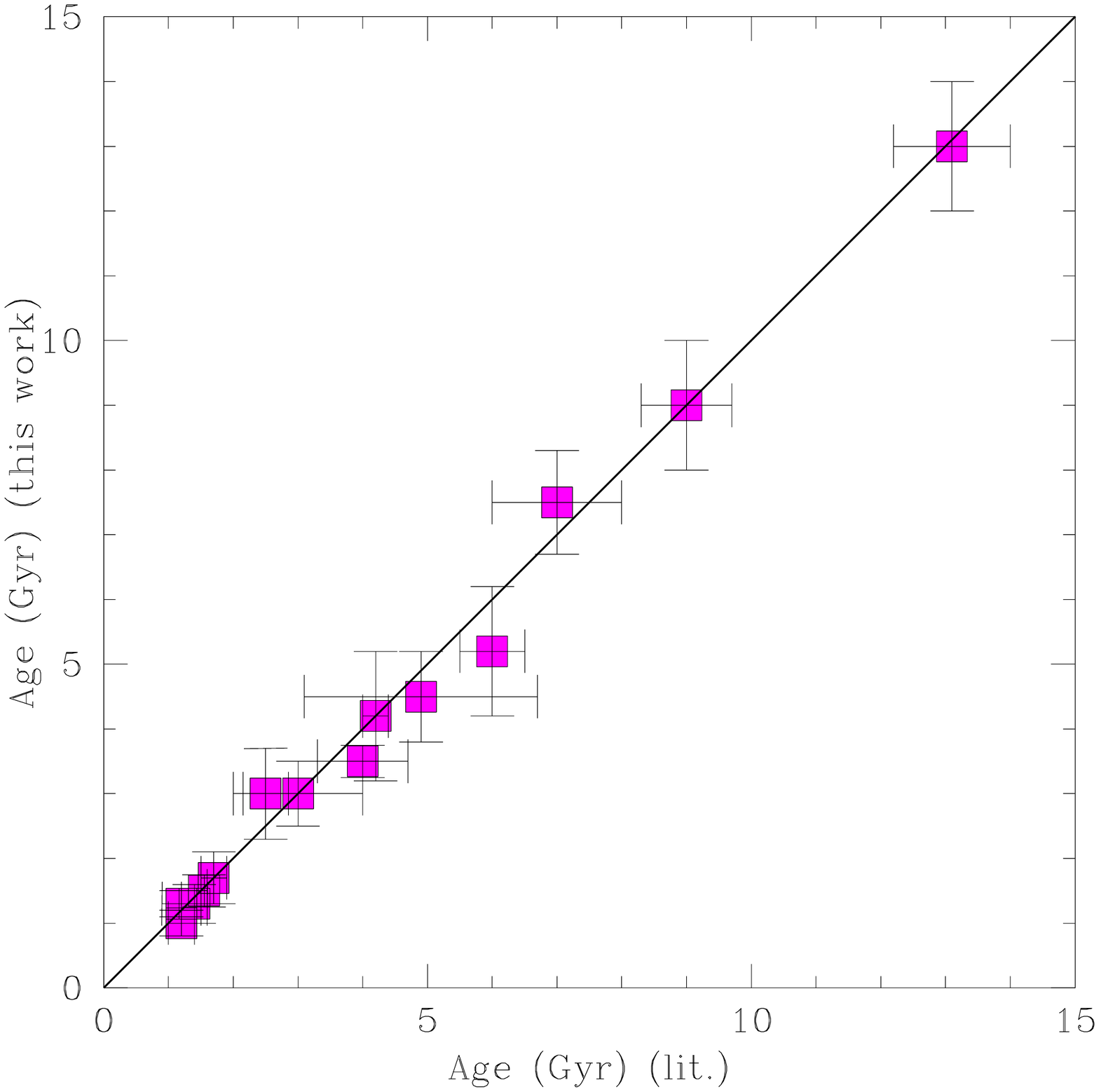}
\caption{Comparison between ages taken from the literature and estimated using the age-metallicity
diagnostic diagram.}
\label{fig6}
\end{figure*}

\begin{figure*}
\includegraphics[width=144mm]{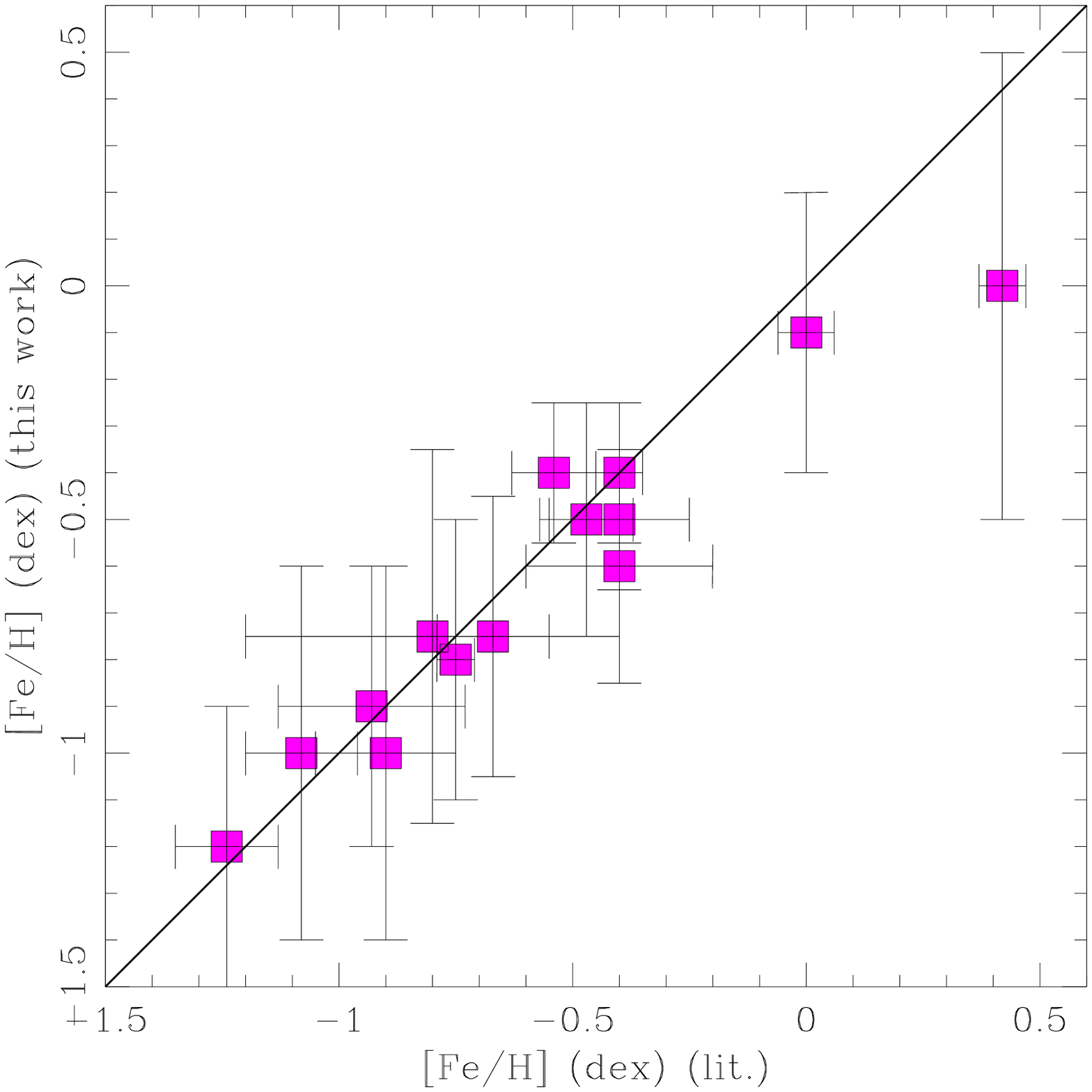}
\caption{Comparison between metallicities taken from the literature and estimated using the age-metallicity
diagnostic diagram.}
\label{fig7}
\end{figure*}



\label{lastpage}
\end{document}